\documentstyle[11pt,aaspp4,amssym]{article}

\received{}
\accepted{}
\journalid{}{}
\articleid{}{}
\slugcomment{Re-Submitted to ApJL 24-Jul-98}
\lefthead{Kippen et al.\ }
\righthead{Gamma-Ray Bursts and Supernovae}

\begin{document}

\title{On the Association of Gamma-Ray Bursts with Supernovae}

\author{R.~M.~Kippen,\altaffilmark{1,2,3}
	M.~S.~Briggs,\altaffilmark{4,2}
	J.~M.~Kommers,\altaffilmark{5}
	C.~Kouveliotou,\altaffilmark{6,2} \\
        K.~Hurley,\altaffilmark{7}
	C.~R.~Robinson,\altaffilmark{6,2}
	J.~van~Paradijs,\altaffilmark{5,8}
	D.~H.~Hartmann,\altaffilmark{9} \\
	T.~J.~Galama,\altaffilmark{8}
	and
	P.~M.~Vreeswijk\altaffilmark{8}}

\altaffiltext{1}{Center for Space Plasma and Aeronomic 
		 Research, University of Alabama in Huntsville, AL 35899}

\altaffiltext{2}{NASA/Marshall Space Flight Center, Huntsville, AL 35812}

\altaffiltext{3}{Mailing address: ES-84, NASA/Marshall Space Flight
                 Center, Huntsville, AL 35812; marc.kippen@msfc.nasa.gov}

\altaffiltext{4}{Physics Department, University of Alabama in
                 Huntsville, AL 35899}

\altaffiltext{5}{Department of Physics and Center for Space Research,
                 Massachusetts Institute of Technology, Cambridge,
		 MA 02139}

\altaffiltext{6}{Universities Space Research Association, Huntsville, AL 35805}

\altaffiltext{7}{Space Sciences Laboratory, University of California, 
		 Berkeley, CA 94720-7450}

\altaffiltext{8}{Astronomical Institute ``Anton Pannekoek'',
                 University of Amsterdam \& Center for High Energy
                 Astrophysics, Kruislaan 403, 1098 SJ Amsterdam, The
                 Netherlands}

\altaffiltext{9}{Department of Physics and Astronomy, Clemson University, 
		 Clemson, SC 29634}


\begin{abstract}

The recent discovery of a supernova (SN~1998bw) seemingly associated
with GRB~980425 adds a new twist to the decades-old debate over the
origin of gamma-ray bursts.  To investigate the possibility that some
(or all) bursts are associated with supernovae, we performed a
systematic search for temporal/angular correlations using catalogs of
BATSE and BATSE/{\it Ulysses\/} burst locations.  We find no
associations with any of the precise BATSE/{\it Ulysses\/} locations,
which allows us to conclude that the fraction of high-fluence
gamma-ray bursts from known supernovae is small ($<$0.2\%).  For the
more numerous weaker bursts, the corresponding limiting fraction of
1.5\% is less constraining due to the imprecise locations of these
events.  This limit ($1.5\% \simeq 18$ bursts) allows that a large
fraction of the recent supernovae used as a comparison data set ($18$
supernovae $\simeq 20\%$) could have associated gamma-ray bursts.
Thus, although we find no significant evidence to support a
burst/supernova association, the possibility cannot be excluded for
weak bursts.

\end{abstract}


\pagebreak

\section{Introduction}

Ever since the discovery of gamma-ray bursts (GRBs) more than 30 years
ago, researchers have sought to associate them with known
astrophysical objects.  This effort was fruitless until recently when
accurate and timely GRB locations provided by the {\it Beppo\/}SAX
Wide Field Cameras led to the discovery of X-ray/optical/radio
transients and associated host galaxies (Costa et al.\ 1997; van Paradijs
et al.\ 1997; Frail et al.\ 1997).  While the objects causing
the bursts, themselves, still remain a mystery, it is now possible to
directly identify and study their host galaxies.  In particular,
optical spectroscopy of two host objects (presumably dim galaxies) has
provided direct measurement of their redshift/distance: for GRB~970508
the redshift $z = 0.835$ (Metzger et al.\ 1997) and for GRB~971214 $z
= 3.42$ (Kulkarni et al.\ 1998).  These measurements provide the first
direct evidence of the cosmological origin of GRBs---something which
was strongly suggested by many years of {\it Compton\/}-BATSE
measurements indicating an isotropic, inhomogeneous spatial
distribution of GRB sources (e.g., Briggs et al.\ 1996; Pendleton et
al.\ 1996).

An apparent discrepancy with the now ``standard'' picture of GRBs
originating in distant galaxies is the recent discovery of a bright
supernova (SN~1998bw) within the 8$\arcmin$ (radius) {\it Beppo\/}SAX
Wide Field Camera error circle of GRB~980425 (Galama et al.\ 1998a).
Galama et al.\ (1998b) estimated, with conservative {\it a
posteriori\/} assumptions, that the chance probability of such an
occurrence is $\sim$10$^{-4}$.  It is thus difficult to reject the
hypothesis that SN~1998bw is related to GRB~980425, even though the
optical lightcurve is markedly different from those of other GRB
optical afterglows.  This conflicts with the standard view of GRB
origin because SN~1998bw resides in a relatively nearby galaxy
(ESO~184-G82) with redshift $z = 0.0085$ (Tinney et al.\ 1998) and the
burst was average in its gamma-ray properties (e.g., peak intensity,
spectrum, fluence, duration; Kippen et al.\ 1998a) compared to other
BATSE bursts.  The notion that an average burst could result from a
nearby supernova event is difficult to reconcile with a population of
bursts originating at cosmological distances.

In this {\it Letter\/} we report on a systematic search for similar
GRB/Supernova coincidences that may have occurred since 1991 April,
when BATSE began observing bursts.  We utilize numerous precise GRB
locations provided by BATSE/{\it Ulysses\/} Interplanetary Network
(IPN) timing analysis, in addition to the less precise (but more
numerous) events localized with BATSE alone.  At the end of the {\it
Letter\/}, we compare our results to those of an independent study
(Wang \& Wheeler 1998) which reported a correlation between GRBs and
type Ib/c supernovae.


\section{Supernova Catalog \& Selection Criteria}

The working hypothesis in this study, based on the possibility that
GRB~980425 and SN~1998bw are associated, is that some fraction of GRBs
may be related to supernova explosions.  We therefore search for
coincidences, both in time and direction, between known GRBs and known
supernovae.  As a comparison dataset, we use a list of all reported
supernovae obtained from the Central Bureau of Astronomical Telegrams
\footnote{http://cfa-www.harvard.edu/iau/lists/Supernovae.html}.  This
list includes the date of discovery, the SN position and/or the host
galaxy position, the optical magnitude at the time of discovery, and
the SN type (if it has been determined).  At faint magnitudes, the
supernova sample is highly biased by deep search campaigns that scan
small regions of the sky.  The bias is less severe for brighter events
that are more easily detected and we therefore begin by limiting the
sample to magnitudes $M \leq 17$ ($\sim$3 mag fainter than SN~1998bw
at the time of its discovery).  We also reject the few SN with
unspecified magnitudes and use the host galaxy position when the SN
position is not specified.  These selections yield a list of 160
supernovae with discovery dates from 1991 Feb.\ through 1998
May---including 58 events classified as SN type Ia, 11 as type Ib/Ic
and 91 others (either type II, or unclassified).  The only other
selection criterion is the choice of a time window $\Delta{T}$ (SN
discovery time minus the GRB time) wherein GRBs will be considered as
possibly related to supernovae.  Given the considerable spread
($\sim$days to weeks) in the time from a SN explosion to its optical
discovery, we allow a generous window of $0 \leq \Delta{T} \leq 30$
days as our baseline selection.

\section{GRB/Supernova Correlation Analysis}

We employ a standard approach for examining the GRB/SN correlation.
When comparing a sample of GRBs to the SN list described above, the
number of GRB/SN pairs within the $\Delta{T}$ search window are
counted to provide a measure of the temporal association.  Directional
association is measured by only including those pairs where the SN
position is within the corresponding GRB error box (we use the term
``error box'' to describe the region of uncertainty around a GRB
location, which can be of arbitrary shape).  The exact definition of
what constitutes a GRB error box depends on the GRB catalog being
considered as described below.  The total number of GRB/SN pairs
$N^{\rm P}$ meeting these criteria is used as a correlation statistic.

To assess the significance of the measured correlation statistic we
employ Monte Carlo simulations wherein 10$^4$ random GRB catalogs are
generated under the null-hypotheses that GRBs are distributed
uniformly in time; distributed isotropically on the sky (corrected for
the observing biases of the particular GRB catalog being considered);
and are uncorrelated with supernovae.  The correlation statistic is
computed between each random catalog and the supernovae list---thereby
providing a measure of the statistical distribution of $N^{\rm P}$
values.  The significance $P$ of the observed correlation $N^{\rm
P}_{\rm obs}$ is given by the fraction of simulated catalogs having
$N^{\rm P} \geq N^{\rm P}_{\rm obs}$.

In the case of GRB catalogs with imprecise locations (where many
chance coincidences are probable), an alternative test is used that
considers the entire distribution of GRB/SN angular deviations, rather
than simply testing if supernovae locations are inside or outside
pre-defined GRB error boxes.  In this test the Kolmogorov-Smirnov
technique (see Press et al.\ 1989) is applied between the cumulative
distribution of the measured GRB/SN angular deviations and that from
the average of many simulated GRB catalogs.  The parameter of interest
is the maximum absolute deviation $D_{\rm KS}$ between the two
distributions.  The significance $P_{\rm KS}$ of a measured
correlation is given by the fraction of simulated catalogs having
$D_{\rm KS} \geq$ the measured value.

\subsection{BATSE/{\it Ulysses\/} Results}

The BATSE 4B (revised) catalog contains locations for 1637 GRBs that
were detected from 1991 April through 1996 Sept.\ (Paciesas et al.\
1998).  Unfortunately, BATSE locations are not particularly useful for
correlation studies due to their large uncertainties (several degrees,
on average).  One way to improve the precision is to combine the BATSE
data with BATSE/{\it Ulysses} (hereafter B/U) IPN timing annuli (see
e.g., Kippen, Hurley \& Pendleton 1998).  Since the annuli are very
precise (typically $\sim$arc-minutes, in one dimension), the
intersection of an annulus with its corresponding BATSE localization
results in an error box typically 25 times smaller in area than the
BATSE localization alone.  IPN annuli are available for the 415 BATSE
4B (revised) bursts also detected by {\it Ulysses\/} (Hurley et al.\
1998a, 1998b; Laros et al.\ 1998).  In general, these are the
$\sim$25\% most fluent BATSE bursts---a reflection of the difference
in sensitivity between {\it Ulysses\/} and BATSE.  IPN data for bursts
beyond the 4B catalog interval are available, but we do not use them
since they have not undergone final processing and checking.

In the correlation analysis, we define a combined B/U error box to be
the intersection of the 3$\sigma$ timing annulus (Hurley et al.\ 1998a)
with the corresponding 99.73\% confidence (i.e., 3$\sigma$) BATSE
error circle (as computed with the ``core-plus-tail'' error
distribution model of Briggs et al.\ 1998a, b; hereafter the CPT
model).  The use of such a large BATSE error circle does not cause
significant loss of correlation sensitivity because the timing annuli
are so precise.  In performing the required Monte Carlo simulations,
burst positions are sampled according to the BATSE 4B exposure
function (Hakkila et al.\ 1998) and then randomly displaced according
to the real B/U burst location uncertainty distributions.  Location
displacement is important, for the uncertainty in a B/U location is
highly correlated with its position on the sky---a result of the fact
the {\it Ulysses\/} spacecraft follows a non-random trajectory.

Comparing the 415 combined B/U burst locations with the SN catalog, we
find 585 GRB/SN pairs within the 30 day search window. In {\it none\/}
of these pairs is the SN location within the GRB error box.  This
measured value of $N^{\rm P}_{\rm obs} = 0$ is consistent with the
number $\langle N^{\rm P}\rangle = 0.06 \pm 0.24$ expected by chance
and nearly independent of the size of the $\Delta{T}$ search
window. (In fact, the only 2 spatially coincident GRB/SN pairs meeting
our criteria have $\Delta{T} = 826$ days and 844 days, respectively).
We performed further simulations wherein a fraction of the bursts
$F_{\rm B}$ were forced to originate (before applying the location
displacement process) at a cataloged SN location.  The value of
$N^{\rm P}_{\rm obs} = 0$ is inconsistent (at the 99.5\% confidence
level) with even a single B/U burst originating at any of the reported
supernovae locations.  We thus have a conservative limit that $F_{\rm
B} < 1/415$.  However, there is the possibility that $F_{\rm B}$ could
be larger for low-fluence bursts not included in the B/U sample.

\subsection{BATSE-only Results}

To investigate the possible correlation of weaker GRBs with
supernovae, we must use the BATSE locations alone.  This significantly
reduces the correlation sensitivity---especially for weak bursts,
which have larger-than-average location uncertainties.  Due to the
imprecise locations, and the large number of BATSE bursts, many chance
GRB/SN coincidences are likely.  The correlation sensitivity is thus a
strong function of the GRB error circle definition used in computing
the $N^{\rm P}$ statistic.  We performed simulations varying the
confidence level of the BATSE error circle and find that using
$\sim$68.27\% confidence (i.e., $\sim$1$\sigma$) BATSE error circles
nearly optimizes the sensitivity, with the ability to detect $F_{\rm
B} \sim 1\%$.  Since we have already found that none of the B/U bursts
are associated with a SN, including them in the BATSE-alone analysis
will only add statistical noise.  We therefore consider only the 1222
BATSE 4B (revised) bursts that are not included in the B/U sample
(referred to as the 4B$-$IPN sample).  For these bursts, $N^{\rm
P}_{\rm obs} = 9$, whereas $\langle N^{\rm P}\rangle = 6.9 \pm 2.6$
are expected by chance.  This is an insignificant excess, found to
occur with a probability $P = 0.260$.  The alternative correlation
test confirms this result, yielding a probability value of $P_{\rm KS}
= 0.650$.  It is possible that other SN selection criteria could yield
different results.  We therefore repeated the analysis, varying the
$\Delta{T}$ temporal window and SN magnitude limit.  As shown in
Figure~1, no significant correlation was found for any selection.  The
lowest probability is at the marginal $\sim$7\% level, which we judge
to be insignificant given the many trials of different selection
parameters.

We are fortunate to have two large, independent samples of weak GRB
locations to examine: BATSE post-4B bursts (Meegan et al.\ 1998) and
untriggered BATSE bursts (Kommers et al.\ 1997, 1998).  As indicated
in Table~1, there are no significant excess correlations for either of
these data sets.  Table~1 also includes results using separate SN
types, for which there are also no significant correlations.  Lacking
any significant detections, we performed additional simulations
varying $F_{\rm B}$.  For the 4B$-$IPN sample, we find that the data
require $F_{\rm B} < 1.5\%$ at the 99.9\% confidence level.

\section{Comparison with Other Results}

Our analysis yields no significant evidence of an association between
GRBs and known supernovae.  This is in contradiction with the results
of an independent study performed by Wang \& Wheeler (1998, hereafter
WW98), where a significant correlation with type Ib/c supernovae was
reported.  To examine whether the two studies are discrepant, we
applied our correlation analysis technique to the list of 21
supernovae used by WW98 and the full ``current'' BATSE catalog, which
includes all bursts detected through 1998 May.  We find 229 GRB/SN
pairs within the supernovae temporal windows specified by WW98.  None
of these pairs has the SN position within the corresponding BATSE
68.27\% confidence GRB error radius---consistent with the number
$\langle N^{\rm P}\rangle = 0.68 \pm 0.85$ expected by chance.  The
alternate correlation technique, which considers the full distribution
of GRB/SN angular separations, yields $P_{\rm KS} = 0.1702$.  If, like
WW98, we limit the sample to only consider the 6 type Ib/c supernovae,
the value becomes $P_{\rm KS} = 0.0946$.  We thus conclude that there
is no significant correlation---in agreement with our original
results.  It appears that the WW98 result is an artifact of their use
of an inappropriate or errant definition of BATSE location errors.  As
indicated in Table~2, the 6 GRB/SN Ib/c associations proposed by WW98
are not very likely based upon the angular separations and the CPT
model of the BATSE error distribution.  In fact, only four of the
proposed associations have separations within the BATSE 99.73\%
confidence (i.e., 3$\sigma$) limits.  If a constant 1.6$\arcdeg$ BATSE
systematic error model is used instead of the more appropriate CPT
model, only three of the pairs are left within the BATSE 3$\sigma$
confidence circles.  It is unclear what BATSE error model WW98 used to
find their 6 proposed associations.  Further evidence against the
reality of the SN Ib/c correlation is given by the fact that 2 of the
6 proposed associations (GRB~960221/SN~1996N and GRB~980218/SN~1998T)
are completely ruled-out by BATSE/{\it Ulysses\/} IPN timing annuli.
The remaining 4 bursts were not detected by {\it Ulysses\/}.

\section{Discussion}

Our study shows that no GRB with a precise BATSE/{\it Ulysses\/}
location is associated with any known supernova (within our search
constraints).  We are thus able to conclude, with high confidence,
that the fraction of high-fluence bursts associated with {\it known\/}
supernovae is small ($F_{\rm B} < 0.2\%$).  The result for weaker
bursts that $F_{\rm B} < 1.5\%$ is far less constraining, due to the
greater number and much larger location uncertainties of these events.
Thus, although we found no significant evidence for a GRB/SN
association, this possibility still exists for weak bursts.  In fact,
the $F_{\rm B} < 1.5\% $ limit still allows that $\sim$20\% of the
reported supernovae we considered could have an associated BATSE GRB.
The incompleteness of the supernova sample makes the result for weak
events even less constraining.  The expected number of supernovae
reaching a {\em peak\/} with $\rm{m_B}$ brighter than 16 is estimated
to be 120--150 yr$^{-1}$ (Galama et al.\ 1998b), whereas the detected
rate for these bright events is about 12 per year from the start of
BATSE.  Therefore, the number of undetected bright events is about
90$\%$ of the total number of supernovae.  Our limits on the fraction
of GRBs from known supernovae must therefore be increased by $\sim$10
times when applied all supernovae.

An independent limit on the fraction of GRBs related to
non-cosmological supernovae is given by the shape of the burst
$V/V_{max}$ distribution.  This distribution should be uniform between
0 and 1 for bursts of such local origin (Schmidt, Higdon, \& Heuter
1988).  In contrast, the measured distribution is highly non-uniform,
with most bursts concentrating at $V/V_{max} < 0.5$.  The smallest
measured values of $V/V_{max}$ thus set an upper limit on the
normalization of any uniformly distributed component in the full
distribution.  The most sensitive search so far for faint GRBs is that
carried out by Kommers et al.\ (1997, 1998), which can detect bursts
that have peak fluxes lower by a factor of 2 than those detected with
the onboard BATSE burst trigger.  In their sample of 2267 GRBs, only
26 have $0.8 < V/V_{max} < 1.0$; so at most only $5 \times (26 \pm 5)
= 130 \pm 25$ of the 2267 bursts could come from a spatially
homogeneous population (Kommers et al.\ 1998).  This corresponds to
just $6 \pm 1$ percent of all bursts that could be associated with
``nearby'' supernovae.

This brings us back to the instigator of this study---the
GRB~980425/SN~1998bw association.  If the association is real and (as
the BATSE data allow, but do not indicate) a small fraction of weak
GRBs are associated with supernovae, we are left to explain why the
association does not also exist for strong bursts.  The hypothesis put
forth by Wang \& Wheeler (1998) is that all GRBs are related to
supernovae, but only for some nearby events (presumably those like
SN~1998bw) are we able observe a weak isotropic emission.  For the
more numerous cosmologically distant sources, the isotropic component
is unobservable, and the bursts are due to highly collimated emission.
The other possibility is that bursts like GRB~980425 are the result of
rare SN events and are completely unrelated to the bulk of gamma-ray
bursts (Woosley, Eastman \& Schmidt 1998; Iwamoto et al.\ 1998).  This
would imply that different mechanisms can produce bursts with similar
gamma-ray properties.  Woosley, Eastman \& Schmidt (1998) suggested
that GRB~970514 may be another example of this phenomena, since the
BATSE location and time of this burst are coincident with
SN~1997cy---a particularly luminous SN with strange spectral
properties.  (Unfortunately, an IPN annulus is not available).  It is
possible that studies of the spectral and temporal properties of GRB
sub-sets could provide some further insight (see e.g., Bloom et al.\
1998).  However, without a large sample of accurate locations for weak
bursts, it is unlikely that either of these scenarios can be ruled
out.

\acknowledgments

This research was supported, in part, through the {\it Compton\/}
Gamma Ray Observatory guest investigator program under grants
NAG5-6747, NAG5-4799 and NAG5-3674.  DHH also acknowledges support
from this program.  JMK acknowledges support from NASA Graduate
Student Researchers Fellowship NGT-52816.


\newpage

\begin{figure}[ht]
\centerline{\epsscale{0.5}\plotone{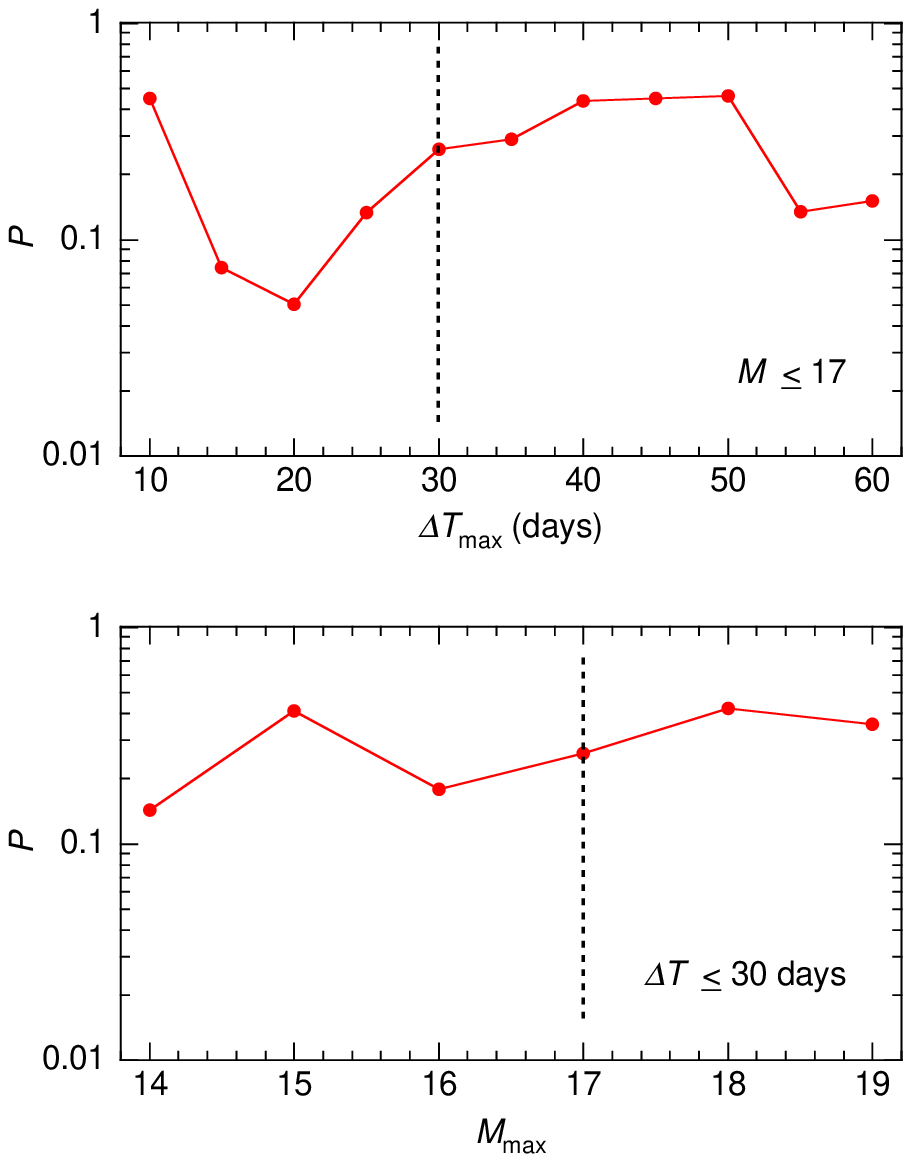}}
\centerline{\parbox[]{14cm}{
\figcaption[f1.eps]{\baselineskip=9pt\footnotesize Gamma-ray burst/
supernova correlation significance $P$ (small number indicates higher
significance) as a function of the supernova magnitude selection $M
\leq M_{\rm max}$ and temporal window $0 \leq \Delta{T} \leq
\Delta{T}_{\rm max}$.  Dashed lines indicate the baseline search
criteria. \label{fig1} } } }
\end{figure}

\newpage

\begin{deluxetable}{lccrrrr}
\tablecolumns{8}
\tablewidth{0pc}
\tablenum{1}
\tablecaption{GRB/SN Correlation Statistics \label{tab1} }
\tablehead{
\colhead{GRB Catalog} & \colhead{No. GRBs} & \colhead{SN Type} &
\colhead{$N^{\rm P}_{\rm obs}$} & \colhead{$\langle N^{\rm P}\rangle$} & 
\colhead{$P$} & \colhead{$P_{\rm KS}$} }
\startdata
B/U IPN           &  415\phn & All &  0\phn & 0.06 $\pm$      0.24 & 1.0000 & \nodata \nl
BATSE 4B$-$IPN    & 1222\phn & All & 12\phn &  6.9 $\pm$ {\phn}2.6 & 0.2598 & 0.6504 \nl
                  &          & Ia  &  1\phn &  2.3 $\pm$ {\phn}1.5 & 0.8967 & 0.7172 \nl
                  &          & Ib/c&  0\phn &  0.4 $\pm$ {\phn}0.6 & 1.0000 & 0.3527 \nl
                  &          & Other& 8\phn &  4.3 $\pm$ {\phn}2.1 & 0.0684 & 0.4197 \nl
BATSE post 4B     &  496\phn & All &  5\phn &  3.9 $\pm$ {\phn}2.0 & 0.3568 & 0.0761 \nl
                  &          & Ia  &  0\phn &  1.8 $\pm$ {\phn}1.4 & 1.0000 & 0.0730 \nl
                  &          & Ib/c&  2\phn &  0.6 $\pm$ {\phn}0.6 & 0.0669 & 0.0840 \nl
                  &          & Other& 3\phn &  1.3 $\pm$ {\phn}1.3 & 0.2492 & 0.8665 \nl
BATSE untriggered &  876\phn & All & 44\phn & 42.7 $\pm$ {\phn}7.2 & 0.4395 & 0.5761 \nl
                  &          & Ia  & 22\phn & 17.0 $\pm$ {\phn}4.4 & 0.1532 & 0.4214 \nl
                  &          & Ib/c&  4\phn &  3.0 $\pm$ {\phn}1.8 & 0.3645 & 0.7526 \nl
                  &          & Other&18\phn & 22.7 $\pm$ {\phn}5.1 & 0.8502 & 0.4963 \nl
\enddata
\tablenotetext{}{Note: These data were obtained using the baseline search
criteria of $M \leq 17$ and $0 \leq \Delta{T} \leq 30$ days.}
\end{deluxetable}

\begin{deluxetable}{lccrccc}
\tablecolumns{7}
\tablewidth{0pt}
\tablenum{2}
\tablecaption{GRB/SN Associations Proposed by WW98 \label{tab2} }
\tablehead{
\colhead{Supernova} & \colhead{GRB} & \colhead{BATSE} &
\colhead{$\Delta{\theta}$\tablenotemark{a}} & 
\colhead{$\sigma_{\rm stat}$\tablenotemark{b}} & 
\colhead{$N\sigma$\tablenotemark{c}} & \colhead{$N\sigma$\tablenotemark{d}} \\
\colhead{} & \colhead{} & \colhead{No.} & \colhead{($\arcdeg$)} & \colhead{($\arcdeg$)} & 
\colhead{} & \colhead{} }
\startdata
1992ad & 920609  &  1641 &  17.5 &  9.99 &  2.0 &  2.1 \nl
1994I  & 940331  &  2900 &  28.6 &  7.89 &  4.4 &  5.0 \nl
1996N  & 960221  &  4959 &   7.8 &  4.01 &  1.8 &  2.3 \nl
1997X  & 970103  &  5740 &  13.4 &  3.87 &  3.1 &  4.5 \nl
1997ei & 971120  &  6488 &  19.7 &  9.94 &  2.4 &  2.5 \nl
1998T  & 980218  &  6605 &   6.3 &  0.93 &  2.0 &  4.8 \nl
\enddata
\tablenotetext{a}{Angular separation between supernova and GRB
location centroid.}
\tablenotetext{b}{BATSE statistical uncertainty radius.}
\tablenotetext{c}{BATSE confidence level at $\Delta{\theta}$
(converted to units of Gaussian standard deviations) based on the
CPT error model.}
\tablenotetext{d}{BATSE confidence level at $\Delta{\theta}$ based on
a 1.6$\arcdeg$ systematic error.}
\end{deluxetable}

\end{document}